\documentclass{ws-procs11x85}

\usepackage{multicol}

\newcommand{\sla}[1]{\,/\!\!\!\!#1}

\newcommand{\BaBar}{\mbox{$\mathrm{Ba\bar{B}ar}$}}
\makeindex

\begin{document}

\title{Statistical Challenges for Searches for New Physics at the LHC}

\author{Kyle Cranmer}

\address{Brookhaven National Laboratory, Upton, NY  11973,  USA\\
e-mail: Kyle.Cranmer@cern.ch}



\twocolumn[\maketitle\abstract{Because the emphasis of the LHC is on
$5\sigma$ discoveries and the LHC environment induces high systematic
errors, many of the common statistical procedures used in High Energy
Physics are not adequate.  I review the basic ingredients of LHC
searches, the sources of systematics, and the performance of several
methods.  Finally, I indicate the methods that seem most promising for
the LHC and areas that are in need of further study.}]

\baselineskip=13.07pt

\section{Introduction}

The Large Hadron Collider (LHC) at CERN and the two multipurpose
detectors, {\sc Atlas} and {\sc CMS}, have been built in order to
discover the Higgs boson, if it exists, and explore the theoretical
landscape beyond the Standard Model.\cite{LHCC99-15,LHCC94-28}  The
LHC will collide protons with unprecedented center-of-mass energy
($\sqrt{s}=14$ TeV) and luminosity ($10^{34}$ cm$^{-2}$s$^{-1}$); the
{\sc Atlas} and {\sc CMS} detectors will record these interactions
with $\sim 10^8$ individual electronic readouts per event.  Because
the emphasis of the physics program is on discovery and the
experimental environment is so complex, the LHC poses new challenges
to our statistical methods -- challenges we must meet with the same
vigor that led to the theoretical and experimental advancements of
the last decade.

In the remainder of this Section, I introduce the physics goals of the
LHC and most pertinent factors that complicate data analysis.  I also
review the formal link and the practical differences between
confidence intervals and hypothesis testing.

In Sec.~\ref{sec:ingredients}, the primary ingredients to new particle
searches are discussed.  Practical and toy examples are presented in
Sec.~\ref{sec:examples}, which will be used to assess the most common
methods in Sec.~\ref{sec:reviewOfMethods}.  The remainder of this
paper is devoted to discussion on the most promising methods for the
LHC.

\subsection{Physics Goals of the LHC}\label{subsec:physicsGoals}

Currently, our best experimentally justified model for fundamental
particles and their interactions is the {\it standard model}.  In
short, the physics goals of the LHC come in two types: those that
improve our understanding of the standard model, and those that go
beyond it.\newline


The only particle of the standard model that has not been observed is
the Higgs boson, which is key for the standard model's description of
the electroweak interactions.  The mass of the Higgs boson, $m_H$, is
a free parameter in the standard model, but there exist direct
experimental lower bounds and more indirect upper bounds.  Once $m_H$
is fixed, the standard model is a completely predictive theory.  There
are numerous particle-level Monte Carlo generators that can be
interfaced with simulations of the detectors to predict the rate and
distribution of all experimental observables.  Because of this
predictive power, searches for the Higgs boson are highly tuned and
often employ multivariate discrimination methods like neural networks,
boosted decision trees, support vector machines, and genetic
programming.\cite{Prosper:2002hw,Friedman:2003ic,cranmer:2005:CPC}

While the Higgs boson is key for understanding the electroweak
interactions, it introduces a new problem: {\it i.e.} the {\it
hierarchy problem}.
There are several proposed solutions to the problem, one of which is
to introduce a new fundamental symmetry, called supersymmetry (SUSY),
between bosons and fermions.
In practice, the minimal supersymmetric extension to the standard
model (MSSM), with its 105 parameters, is not so much a theory as a
theoretical framework.


They key difference between SUSY and Higgs searches is that, in most
cases, discovering SUSY will not be the difficult part.  Searches for
SUSY often rely on robust signatures that will show a deviation from
the standard model for most regions of the SUSY parameter space.  It
will be much more challenging to demonstrate that the deviation from
the standard model is SUSY and to measure the fundamental parameters
of the theory.\cite{Hinchliffe:1996iu} In order to restrict the scope
of these proceedings, I shall focus LHC Higgs searches, where the
issues of hypothesis testing are more relevant.

\subsection{The Challenges of LHC Environment}

The challenges of the LHC environment are manifold.  The first and
most obvious challenge is due to the enormous rate of uninteresting
background events from QCD processes.  The total interaction rate for
the LHC is of order $10^9$ interactions per second; the rate of Higgs
production is about ten orders of magnitude smaller.  Thus, to
understand the background of a Higgs search, one must understand the
extreme tails of the QCD processes.

Compounding the difficulties due to the extreme rate is the complexity
of the detectors.  The full-fledged simulation of the detectors is
extremely computationally intensive, with samples of $10^7$ events
taking about a month to produce with computing resources distributed around
the globe.  This computational limitation constrains the problems that
can been addressed with Monte Carlo techniques.

Theoretical uncertainties also contribute to the challenge.  The
background to many searches requires calculations at, or just beyond,
the state-of-the-art in particle physics.  The most common situation
requires a final state with several well-separated high transverse
momentum objects ({\it e.g.} $t\bar{t}\,jj\to b l \nu \,\bar{b} jj\,j
j $), in which the regions of physical interest are not reliably
described by leading-order perturbative calculations (due to infra-red
and collinear divergences), are too complex for the requisite
next-to-next-to-leading order calculations, and are not properly
described by the parton-shower models alone.  Enormous effort has gone
into improving the situation with next-to-leading order calculations
and matrix-element--parton-shower
matching.\cite{Frixione:2002bd,Schalicke:2005nv} While these new tools
are a vast improvement, the residual uncertainties are still often
dominant.

Uncertainties from non-perturbative effects are also important.  For
some processes, the relevant regions of the parton distribution
functions are not well-measured (and probably will not be in the first
few years of LHC running), which lead to uncertainties in rate as well
as the shape of distributions.  Furthermore, the various
underlying-event and multiple-interaction models used to describe data
from previous colliders show large deviations when extrapolated to the
LHC.\cite{Moraes}  This soft physics has a large impact on the
performance of observables such as missing transverse energy.

In order to augment the simulated data chain, most searches introduce
auxiliary measurements to estimate their backgrounds from the data
itself.  In some cases, the background estimation is a simple
sideband, but in others the link between the auxiliary measurement to
the quantity of interest is based on simulation.  This hybrid approach
is of particular importance at the LHC.

While many of the issues discussed above are not unique to the LHC,
they are often more severe.  At LEP, it was possible to generate Monte
Carlo samples of larger size than the collected data, QCD backgrounds
were more tame, and most searches were not systematics-limited.  The
Tevatron has much more in common with the LHC; however, at this point
discovery is less likely, and most of the emphasis is on measurements
and limit setting.

\subsection{Confidence Intervals \& Hypothesis Testing}\label{subsec:IntervalsVsTesting}

The last several conferences in the vein of {\it PhyStat 2005} have
concentrated heavily on confidence intervals.  In particular, 95\%
confidence intervals for some physics parameter in an experiment that
typically has few events.  More recently, there has been a large
effort in understanding how to include systematic errors and nuisance
parameters into these calculations.  

LHC searches, in contrast, are primarily interested in $5\sigma$
discovery.  The $5\sigma$ discovery criterion is somewhat vague, but
usually interpreted in a frequentist sense as a hypothesis test with
a rate of Type~I error $\alpha = 2.85 \cdot 10^{-7}$.

There is a formal link between confidence intervals and hypothesis
testing: frequentist confidence intervals from the Neyman construction
are formally inverted hypothesis tests.  It is this equivalence that
links the Neyman-Pearson lemma%
\footnote{The lemma states that, for a simple hypothesis test of size
$\alpha$ between a null $H_0$ and an alternate $H_1$, the most powerful
critical region in the observable $x$ is given by a contour of the
likelihood ratio $L(x|H_0)/L(x|H_1)$.} %
to the ordering rule used in the {\it unified method} of Feldman and
Cousins.\cite{Feldman:1998qc}  Furthermore, this equivalence will be
very useful in translating our understanding of confidence intervals to
the searches at the LHC.

In some cases, this formal link can be misleading.  In particular,
there is not always a continuous parameter that links the fully
specified null hypothesis $H_0$ to the fully specified alternate $H_1$
in any physically interesting or justified way.  Furthermore, the
performance of a method for a 95\% confidence interval and a $5\sigma$
discovery can be quite different.

%
%
%
%

\section{The Ingredients of an LHC Search}\label{sec:ingredients}

In order to assess the statistical methods that are available and
develop new ones suited for the LHC, it is necessary to be familiar
with the basic ingredients of the search.  In this section, the basic
ingredients, terminology, and nomenclature are established.

\begin{figure}
\epsfig{file=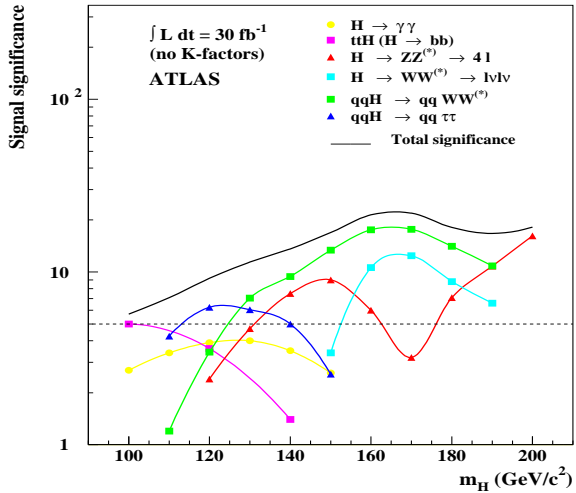, width=.47\textwidth, height=.4\textwidth}
\caption{Expected significance as a function of Higgs mass for the
Atlas detector with 30 fb$^{-1}$ of data.}\label{fig:combination}
\end{figure}

\subsection{Multiple Channels \& Processes}\label{S:MultipleChannels}

Almost all new particle searches do not observe the particle directly,
but through the signatures left by the decay products of the particle.
For instance, the Higgs boson will decay long before it interacts with
the detector, but its decay products will be detected.  In many cases,
the particle can be produced and decay in many different
configurations, each of which is called a {\it search channel} (see
Tab.~\ref{tab:smHiggs}).  There are may be multiple signal and
background processes which contribute to each channel.  For example,
in $H\to \gamma \gamma$, the signal could come from any Higgs
production mechanism and the background from either continuum
$\gamma\gamma$ production or QCD backgrounds where jets fake photons.
Each of these processes have their own rates, distributions for
observables, and uncertainties.  Furthermore, the uncertainties between
processes may be correlated.

In general the theoretical model for a new particle has some free
parameters.  In the case of the standard model Higgs, only the mass
$m_H$ is unknown.  For SUSY scenarios, the Higgs model is parametrized
by two parameters: $m_A$ and $\tan\beta$.  Typically, the unknown
variables are scanned and a hypothesis test is performed for each
value of these parameters.  The results from each of the search
channels can be combined to enhance the power of the search, but one
must take care of correlations among channels and ensure consistency.  

The fact that one scans over the parameters and performs many
hypothesis tests increases the chance that one finds at least one
large fluctuation from the null-hypothesis.  Some approaches
incorporate the number of trials explicitly,\cite{Abbott:2000fb}
some approaches only focus on the most interesting
fluctuation,\cite{Gao:2005it} and some see this heightened rate of
Type I error as the motivation for the stringent $5\sigma$
requirement.\cite{Feldman:phystat2005}

\subsection{Discriminating Variables \& Test Statistics}\label{S:DiscriminatingVariables}

Typically, new particles are known to decay with certain
characteristics that distinguish the {\it signal} events from those
produced by {\it background} processes.  Much of the work of a search
is to identify those observables and to construct new discriminating
variables (generically denoted as $m$).  Examples include angles
between particles, invariant masses, and particle identification
criterion.  Discriminating variables are used in two different ways:
to define a signal-like region and to weight events.

The usage of discriminating variables is related to the test
statistic: the real-valued quantity used to summarize the experiment.
The test statistic is thought of as being ordered such that either
large or small values indicate growing disagreement with the null
hypothesis.

A simple ``cut analysis'' consists of defining a signal-like region
bounded by upper- and lower-values of these discriminating variables
and counting events in that region.  In that case, the test statistic
is simply the number of events observed in the signal like region.
One expects $b$ background events and $s$ signal events, so the
experimental sensitivity is optimized by adjusting the cut values.
More sophisticated techniques use multivariate algorithms, such as
neural networks, to define more complicated signal-like regions, but
the test statistic remains unchanged.  In these number counting
analyses, the likelihood of observing $n$ events is simply given by
the Poisson model.

There are extensions to this number-counting technique.  In
particular, if one knows the distribution of the discriminating
variable $m$ for the background-only (null) hypothesis, $f_b(m)$, and
the signal-plus-background (alternate) hypothesis, $f_{s+b}(m) =
[sf_s(m) + bf_b(m)] / (s+b)$, then there is a more powerful test
statistic than simply counting events.  This is intuitive, a well
measured 'golden event' is often more convincing than a few messy
ones.  Following the Neyman-Pearson lemma, the most powerful test
statistic is
\begin{eqnarray}
Q &=& \frac{L({\bf m}|H_1)}{L({\bf m}|H_0) } \\\nonumber
&=& 
\frac{ 
\prod_i^{N_{chan}} Pois(n_i | s_i + b_i)  \prod_j^{n_i} \frac{s_i { f_s(m_{ij})} + b_i { f_b(m_{ij})} }{s_i + b_i}  }
{\prod_i^{N_{chan}} Pois(n_i | b_i)  \prod_j^{n_i} { f_b(m_{ij})} } 
\end{eqnarray}
($n_i$ denotes events in $i^{th}$ channel) or equivalently
\begin{equation}\label{eq:testStatistic}
q = \ln Q = {-s_{tot}}  + \sum_i^{N_{chan}} \sum_j^{n_i} \ln \left( 1+\frac{s_i { f_s(m_{ij})}}{b_i { f_b(m_{ij})}}\right).
\end{equation}

The test statistic in Eq.~\ref{eq:testStatistic} was used by the LEP
Higgs Working Group (LHWG) in their final results on the search for
the Standard Model Higgs.\cite{Barate:2003sz}

At this point, there are two loose ends: how does one determine the
distribution of the discriminating variables $f(m)$, and how does one
go from Eq.~\ref{eq:testStatistic} to the distribution of $q$ for
$H_0$ and $H_1$.  These are the topics of the next subsections.

\subsection{Parametric and Non-Parametric Methods}\label{sec:keys}

\begin{figure}
\center
\epsfig{file=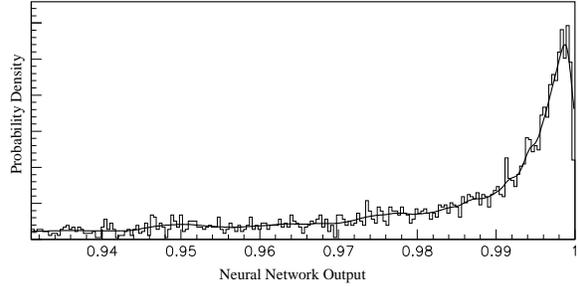,width=.45\textwidth}
\caption{The distribution of a neural network output for signal
events.  The histogram is shown together with $\hat{f}_1(m)$.}
\label{fig:keys}
\end{figure}

In some cases, the distribution of a discriminating variable $f(m)$
can be parametrized and this parametrization can be justified either
by physics arguments or by goodness-of-fit.  However, there are many
cases in which $f(m)$ has a complicated shape not easily parametrized.
For instance, Fig.~\ref{fig:keys} shows the distribution of a neural
network output for signal events.  In that case kernel estimation
techniques can be used to estimate $f(m)$ in a non-parametric way from
a sample of events $\{m_i\}$.\cite{Cranmer:2000du}  The technique
that was used by the LHWG\cite{Barate:2003sz} was based on an adaptive kernel
estimation given by:
\begin{equation}\label{eq:adaptiveKEYS}
\hat{f}_1(m) = \sum_i^n \frac{ 1 }{n h(m_i)} K\left(\frac{m-m_i}{h(m_i)}\right),
\end{equation}
where
\begin{equation}\label{eq:hadaptive}
h(m_i) = \left( \frac{4}{3} \right)^{1/5} \sqrt{\frac{\sigma}{\hat{f}_0(m_i)}} n^{-1/5},
\end{equation}
$\sigma$ is the standard deviation of $\{x_i\}$, $K(x)$ is some kernel
function (usually the normal distribution), and $\hat{f}_0(x)$ is the
fixed kernel estimate given by the same equation but with a fixed $h(m_i)$
\begin{equation}
h^* = \left( \frac{4}{3} \right)^{1/5} \sigma n^{-1/5}.
\end{equation}

The solid line in Fig.~\ref{fig:keys} shows that the method (with
modified-boundary kernels) works very well for shapes with complicated
structure at many scales.

\subsection{Numerical Evaluation of Significance}\label{sec:clfft}

Given, $f_s(m)$ and $f_b(m)$ the distribution of~$q(x)$ can be
constructed.  For the background-only hypothesis, $f_b(m)$ provides
the probability of corresponding values of $q$ needed to define the
single-event pdf $\rho_1$.\footnote{The integral is necessary because
the map $q(m): m \rightarrow q$ may be many-to-one.}
\begin{equation}\label{E:rho_1}
\rho_{1,b}(q_0) = \int f_b(m) \,\delta ( q(m) - q_0 ) dm
\end{equation} 

For multiple events, the distribution of the log-likelihood ratio must
be obtained from repeated convolutions of the single event
distribution.  This convolution can either be performed implicitly
with approximate Monte Carlo techniques,\cite{Junk:1999kv} or
analytically with a Fourier transform technique.\cite{clfft}  In the
Fourier domain, denoted with a bar, the distribution of the
log-likelihood for $n$ events is
\begin{equation}\label{E:rho_n}
\overline{\rho_n} = \overline{\rho_1}^n
\end{equation}
Thus the expected log-likelihood distribution for background with
Poisson fluctuations in the number of events takes the form
\begin{equation}\label{E:rho_b}
  \rho_{b}(q) = \sum_{n=0}^{\infty} \frac{e^{-b} b^n}{n!}\rho_{n,b}(q)
\end{equation}
which in the Fourier domain is simply
\begin{equation}\label{E:exponentiation}
  \overline{\rho_{b}(q)} = e^{b[\overline{\rho_{1,b}(q)} - 1]}.
\end{equation}
For the signal-plus-background hypothesis we expect $s$ events from
the $\rho_{1,s}$ distribution and $b$ events from the $\rho_{1,b}$
distribution, which leads to the expression for $\rho_{s+b}$ in the
Fourier domain\footnote{Perhaps it is worth noting that
$\overline{\rho(q)}$ is a complex valued function of the Fourier
conjugate variable of $q$.  Thus numerically the exponentiation in
Eq.~\ref{E:exponentiation} requires Euler's formula $e^{i\theta} =
\cos \theta + i \sin \theta$.}
\begin{equation}\label{E:sb_exponentiation}
  \overline{\rho_{s+b}(q)} = e^{b[\overline{\rho_{1,b}(q)} - 1] 
                            + s[\overline{\rho_{1,s}(q)} - 1]}.
\end{equation}
This equation generalizes, in a somewhat obvious way, to include many
processes and channels.

Numerically these computations are carried out with the Fast Fourier
Transform (FFT).  The FFT is performed on a finite and discrete array,
beyond which the function is considered to be periodic.  Thus the
range of the $\rho_1$ distributions must be sufficiently large to hold
the resulting $\rho_b$ and $\rho_{s+b}$ distributions.  If they are
not, the ``spill over'' beyond the maximum log-likelihood ratio
$q_{max}$ will ``wrap around'' leading to unphysical $\rho$
distributions.  Because the range of $\rho_b$ is much larger than
$\rho_{1,b}$ it requires a very large number of samples to describe
both distributions simultaneously.  The implementation of this method
requires some approximate asymptotic techniques that describe the
scaling from $\rho_{1,b}$ to $\rho_b$.\cite{physics_03_12050}

The nature of the FFT results in a number of round-off errors and
limit the numerical precision to about $10^{-16}$ -- which limit the
method to significance levels below about $8\sigma$.  Extrapolation
techniques and arbitrary precision calculations can overcome these
difficulties,\cite{physics_03_12050} but such small $p$-values are of
little practical interest.

From the log-likelihood distribution of the two hypotheses we can
calculate a number of useful quantities. Given some experiment with an
observed log-likelihood ratio, $q^*$, we can calculate the
background-only confidence level, $CL_b$ :
\begin{equation}\label{E:clb}
CL_b (q^*) =\int_{q^*}^\infty \rho_b(q')dq'
\end{equation}
In the absence of an observation we can calculate the expected $CL_b$
given the signal-plus-background hypothesis is true. To do this we
first must find the median of the signal-plus-background distribution
$\overline{q}_{s+b}$.  From these we can calculate the expected $CL_b$
by using Eq.~\ref{E:clb} evaluated at $q^* =\overline{q}_{s+b}$.

Finally, we can convert the expected background confidence level into
an expected Gaussian significance, $Z\sigma$, by finding the value of
$Z$ which satisfies
\begin{equation}\label{eq:zvalue}
CL_b(\overline{q}_{s+b}) = \frac{1-{\rm erf}(Z/\sqrt{2})}{2}.
\end{equation}
where ${\rm erf}(Z) = (2/\pi) \int_0^Z \exp( -y^2 )dy$ is a function
readily available in most numerical libraries.  For $Z>1.5$, the
relationship can be approximated\cite{Linnemann} as
\begin{equation}
 Z \approx \sqrt{u-\ln u} \hspace{.1in} {\rm with} \hspace{.1in}  u = -2\ln(CL_b \sqrt{2\pi})
\end{equation}

\subsection{Systematic Errors, Nuisance Parameters \& Auxiliary Measurements}

Sections~\ref{sec:keys} and~\ref{sec:clfft} represent the state of the
art for HEP in frequentist hypothesis testing in the absence of
uncertainties on rates and shapes of distributions.  In practice, the
true rate of background is not known exactly, and the shapes of
distributions are sensitive to experimental quantities, such as
calibration coefficients and particle identification efficiencies
(which are also not known exactly).  What one would call a {\it
systematic error} in HEP, usually corresponds to what a statistician
would refer to as a {\it nuisance parameter}.

Dealing with nuisance parameters in searches is not a new problem, but
perhaps it has never been as essential as it is for the LHC.  In these
proceedings, Cousins reviews the different approaches to nuisance
parameters in HEP and the professional statistical
literature.\cite{Cousins:phystat05} Also of interest is the
classification of systematic errors provided by
Sinervo.\cite{Sinervo:2003wm} In Sec.~\ref{sec:reviewOfMethods}, the a
few techniques for incorporating nuisance parameters are reviewed.

From an experimental point of view, the missing ingredient is some set
of auxiliary measurements that will constrain the value of the
nuisance parameters.  The most common example would be a sideband
measurement to fix the background rate, or some control sample used to
assess particle identification efficiency.  Previously, I used the variable $M$ to
denote this auxiliary measurement\cite{physics_03_10108}; while
Linnemann,\cite{Linnemann} Cousins,\cite{Cousins:phystat05} and Rolke,
Lopez, and Conrad\cite{Rolke:2000ij,Rolke:2004mj} used $y$.
Additionally, one needs to know the likelihood function that provides
the connection between the nuisance parameter(s) and the auxiliary
measurements.

The most common choices for the likelihood of the auxiliary
measurement are $L(y|b) = Pois(y|\tau b)$ and $L(y|b) = G(y|\tau b,
\sigma_y)$, where $\tau$ is a constant that specifies the ratio of the
number of events one expects in the sideband region to the number
expected in the signal-like region.\footnote{Note that
Linnemann\cite{Linnemann} used $\alpha=1/\tau$ instead, but in this
paper $\alpha$ is reserved for the rate of Type~I error.}  %

\begin{figure}
\epsfig{file=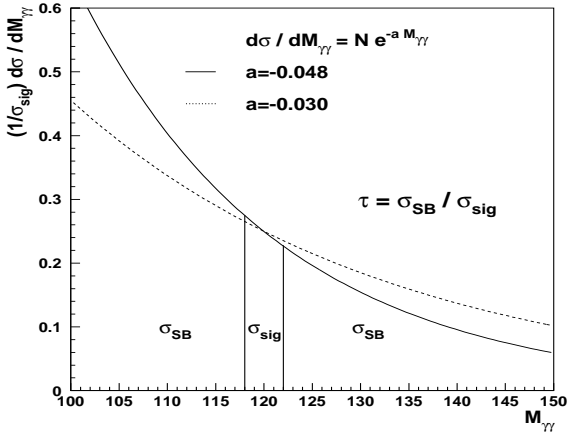,width=.45\textwidth, height=.35\textwidth}
\caption{The signal-like region and sideband for $H\to\gamma\gamma$ in
which $\tau$ is correlated to $b$ via the model parameter $a$.}
\label{fig:hgg_defs}
\end{figure}

A constant $\tau$ is appropriate when one simply counts the number of
events $y$ in an ``off-source'' measurement.  In a more typical case,
one uses the distribution of some other variable, call it $m$, to
estimate the number of background events inside a range of $m$ (see
Fig.~\ref{fig:hgg_defs}).  In special cases the ratio $\tau$ is
independent of the model parameters.  However, in many cases ({\it
e.g.} $f(m) \propto e^{-am}$), the ratio $\tau$ depends on the model
parameters.  Moreover, sometimes the sideband is contaminated with
signal events, thus the background and signal estimates can be
correlated.  These complications are not a problem as long as they are
incorporated into the likelihood.

The number of nuisance parameters and auxiliary measurements can grow
quite large.  For instance, the standard practice at \BaBar\ is to
form very large likelihood functions that incorporate everything from
the parameters of the unitarity triangle to branching fractions and
detector response.  These likelihoods are typically factorized into
multiple pieces, which are studied independently at first and later
combined to assess correlations.  The factorization of the likelihood
and the number of nuisance parameters included impact the difficulty
of implementing the various scenarios considered below.

\section{Practical and Toy Examples}\label{sec:examples}

In this Section, a few practical and toy examples are introduced.  The
toy examples are meant to provide simple scenarios where results for
different methods can be easily obtained in order to expedite their
comparison.  The practical examples are meant to exclude methods that
provide nice solutions to the toy examples, but do not generalize to
the realistic situation.

\subsection{The Canonical Example}

Consider a number-counting experiment that measures $x$ events in the
signal-like region and $y$ events in some sideband.  For a given
background rate $b$ in the signal-like region, say one can expect
$\tau b$ events in the sideband.  Additionally, let the rate of signal
events in the signal-like regions -- the parameter of interest -- be
denoted $\mu$.  The corresponding likelihood function is
\begin{equation}\label{eq:lp}
L_P(x,y|\mu,b) = Pois(x|\mu+b)\cdot Pois(y|\tau b).
\end{equation}
This is the same case that was considered in
Refs.~\cite{Cousins:phystat05,physics_03_10108,Rolke:2000ij,Rolke:2004mj}
for $x,y = \mathcal{O}(10)$ and $\alpha = 5\%$.  For LHC searches, we
will be more interested in $x,y = \mathcal{O}(100)$ and $\alpha = 2.85
\cdot 10^{-7}$.  Furthermore, the auxiliary measurement will rarely be
a pure number counting sideband measurement, but instead the result of
some fit.  So let us also consider the likelihood function
\begin{equation}\label{eq:lg}
L_G(x,y|\mu,b) = Pois(x|\mu+b)\cdot G(y|\tau b, \sqrt{\tau b}).
\end{equation}

As a concrete example in the remaining sections, let us consider the
case $b=100$ and $\tau=1$.  Operationally, one would measure $y$ and
then find the value $x_{crit}(y)$ necessary for discovery.  In the
language of confidence intervals, $x_{crit}(y)$ is the value of $x$
necessary for the $100(1-\alpha)\%$ confidence interval in $\mu$ to
exclude $\mu_0=0$.  In Sec.~\ref{sec:reviewOfMethods} we check the
coverage (Type~I error or false-discovery rate) for both $L_P$ and
$L_G$.

Linnemann reviewed thirteen methods and eleven published examples of
this scenario.\cite{Linnemann}  Of the published examples, only
three (the one from his reference 18 and the two from 19) are near the
range of $x$,$y$, and $\alpha$ relevant for LHC searches.  Linnemann's
review asks an equivalent question posed in this paper, but in a
different way: what is the significance ($Z$ in Eq.~\ref{eq:zvalue})
of a given observation $x,y$.

\subsection{The LHC Standard Model Higgs Search}

\begin{table*}
\caption{Number of signal and background events for representative
Higgs search channels for two values of Higgs mass, $m_H$, with 30 fb$^{-1}$ of
data
.  A rough uncertainty on the
background rate is denoted as $\delta b/b$, without reference to the
type of systematic uncertainty. The table also indicates if the
channels are expected to use a weight $f(m)$ as in
Eq.~\ref{eq:testStatistic}.}
\label{tab:smHiggs} 
\center
\begin{tabular}{|c|rrrccc|}\hline
channel & $s$ & $b$ & $\delta b/b$ & dominant backgrounds & use $f(m)$ & $m_H$ (GeV) \\ \hline
$t\bar{t}H\to t\bar{t}bb$ & 42 & 219 & $\sim$10\% & $t\bar{t}jj, t\bar{t}bb$ & Yes & 120\\
$H\to \gamma\gamma$ & 357 & 11820 & $\sim 0.1$\% & $\gamma\gamma, j\gamma, jj$ & No & 120\\
$qq H\to qq\tau \tau \to qq l l \sla{E}_T$ & 17 & 14  & $\sim$10\% & $Z\to \tau \tau$, $t\bar{t}$ & Yes & 120\\
$qq H\to qq\tau \tau \to qq l h \sla{E}_T$ & 16 & 8  & $\sim$10\% & $Z\to \tau \tau$, $t\bar{t}$ & Yes & 120\\
$qq H\to qqWW^* \to qql l \sla{E}_T$ & 28.5 & 47.4  & $\sim$10\% & $t\bar{t}, WW$ & Yes & 120\\\hline
$qq H\to qqWW^* \to qql l \sla{E}_T$ & 262.5 & 89.1  & $\sim$10\% & $t\bar{t}, WW$ & Yes & 170\\
$H\to ZZ \to 4l$ & 7.6 & 3.1 & $\sim 1$\% & $ZZ\to 4l$ & No & 170\\
$H\to WW \to ll\sla{E}_T$ & 337 & 484 & $\sim$5\% & $Z\to \tau \tau$, $t\bar{t}$ & Yes & 170\\
\hline
\end{tabular}
\end{table*}

The search for the standard model Higgs boson is by no means the only
interesting search to be performed at the LHC, but it is one of the
most studied and offers a particularly challenging set of channels to
combine with a single method.  Figure~\ref{fig:combination} shows the
expected significance versus the Higgs mass, $m_H$, for several
channels individually and in combination for the {\sc Atlas}
experiment.\cite{VBFSCNOTE} Two mass points are considered in more
detail in Tab.~\ref{tab:smHiggs}, including results from
Refs.\cite{LHCC99-15,VBFSCNOTE,ATL-PHYS-2003-024}.  Some of these
channels will most likely use a discriminating variable distribution,
$f(m)$, to improve the sensitivity as described in
Sec.~\ref{sec:keys}.  I have indicated the channels that I suspect
will use this technique.  Rough estimates on the uncertainty in the
background rate have also been tabulated, without regard to the
classification proposed by Sinervo.

\begin{figure}
\epsfig{file=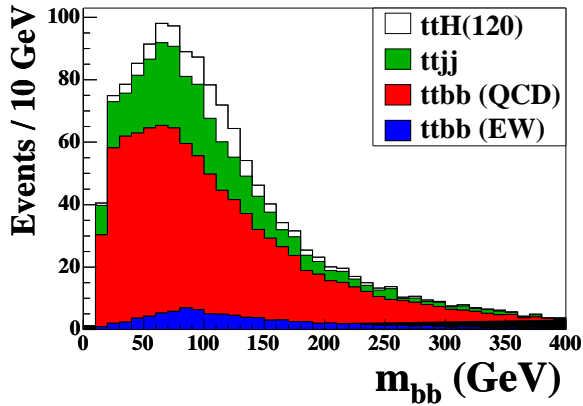, width=.45\textwidth}
\caption{The ${bb}$ invariant mass spectrum for $t\bar{t}H$ signal and
background processes at {\sc
Atlas}.
}\label{fig:tth}
\end{figure}

The background uncertainties for the $t\bar{t}H$ channel have been
studied in some detail and separated into various
sources.\cite{ATL-PHYS-2003-024} Figure~\ref{fig:tth} shows the
$m_{bb}$ mass spectrum for this channel.\footnote{It is not clear if
this result is in agreement with the equivalent {\sc CMS}
result.\cite{CMS-2001-054}} Clearly, the shape of the background-only
distribution is quite similar to the shape of the
signal-plus-background distribution.  Furthermore, theoretical
uncertainties and $b$-tagging uncertainties affect the shape of the
background-only spectrum.  In this case the incorporation of
systematic error on the background rate most likely precludes the
expected significance of this channel from ever reaching $5\sigma$.

\begin{figure}
\epsfig{file=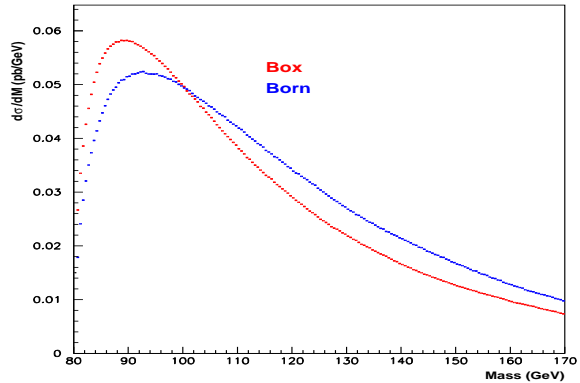, width=.45\textwidth, height=.3\textwidth}
\caption{Two plausible shapes for the continuum ${\gamma\gamma}$
mass spectrum at the LHC.}\label{fig:hgg_spectrum}
\end{figure}

Similarly, the $H\to\gamma\gamma$ channel has uncertainty in the shape
of the $m_{\gamma\gamma}$ spectrum from background processes.  One
contribution to this uncertainty comes from the electromagnetic
energy scale of the calorimeter (an experimental nuisance parameter),
while another contribution comes from the theoretical uncertainty in
the continuum $\gamma\gamma$ production.
Figure~\ref{fig:hgg_spectrum} shows two plausible shapes for the
$m_{\gamma\gamma}$ spectrum from ``Born'' and ``Box'' predictions.

\section{Review of Methods}\label{sec:reviewOfMethods}

Based on the practical example of the standard model Higgs search at
the LHC and the discussion in Sec.~\ref{sec:ingredients}, the list of
admissible methods is quite short.  Of the thirteen methods reviewed
by Linnemann, only five are considered as reasonable or recommended.
These can be divided into three classes: hybrid Bayesian-frequentist methods%
, methods based on the Likelihood Principle, and frequentist methods
based on the Neyman construction.

\subsection{Hybrid Bayesian-Frequentist Methods}\label{sec:CousinsHighland}

The class of methods frequently used in HEP and commonly referred to
as the Cousins-Highland technique (or secondarily Bayes in statistical
literature) are based on a Bayesian average of frequentist $p$-values
as found in the first equation of Ref.\cite{Cousins:1992qz}.  The
Bayesian average is over the nuisance parameters and weighted by the
posterior $P(b|y)$.  Thus the $p$-value of the observation $(x_0,y_0)$
evaluated at $\mu$ is given by

\begin{eqnarray}
p(x_0,y_0|\mu) &=& \int_{0}^\infty db \,p(x_0|\mu,b) P(b|y_0) \label{eq:ch}\\
&=&  \int_{x_0}^\infty dx \,P(x|\mu,y_0) \label{eq:chLEP}
\end{eqnarray}  
where
\begin{equation}\label{eq:chBayes}
P(x|\mu,y_0) = \int_0^\infty db \, P(x|\mu,b)
\frac{P(y_0|b)~P(b)}{P(y_0)}
\end{equation}  
The form in Eq.~\ref{eq:ch}, an average over $p$-values, is similar to
the form written in Cousins \& Highland's article; and it is re-written in
Eq.~\ref{eq:chLEP} to the form that is more familiar to those from LEP
Higgs searches.\cite{Junk:1999kv,clfft}  Actually, the dependence on
$y_0$ and the Bayesian prior $P(b)$ shown explicitly in
Eq.~\ref{eq:chBayes} is often not appreciated by those that use this
method.


The specific methods that Linnemann considers correspond to different
choices of Bayesian priors.  The most common in HEP is to ignore the
prior and use a truncated Gaussian for the posterior $P(b|y_0)$, which
Linnemann calls $Z_N$.  For the case in which the likelihood $L(y|b)$
is known to be Poisson, Linneman prefers to use a flat prior, which
gives rise to a Gamma-distributed posterior and Linnemann's second
preferred method $Z_\Gamma$, which is identical to the ratio of
Poisson means $Z_{Bi}$ and can be written in terms of (in)complete
beta functions as\cite{Linnemann}
\begin{equation}
Z_\Gamma = Z_{Bi} = B(1/(1+\tau), x, y+1)/B(x,y+1).
\end{equation}
The method Linnemann calls $Z_{5'}$ can be seen as an approximation of
$Z_N$ for large signals and is what {\sc Atlas} used to assess its
physics potential.\cite{LHCC99-15} The method not recommended by
Linnemann and was critically reviewed in Ref.\cite{UW-confidence-VBF}.
\begin{equation}
x_{crit}^{5'}(y)=y/\tau + Z \sqrt{y/\tau (1+1/\tau)}
\end{equation}

\subsection{Likelihood Intervals}\label{sec:likelihoodintervals}

As Cousins points out, the professional statistics literature seems
less concerned with providing correct coverage by construction, in
favor of likelihood-based and Bayesian methods.  The likelihood
principle states that given a measurement $x$ all inference about
$\mu$ should be based on the likelihood function $L(x|\mu)$.  When
nuisance parameters are included, things get considerably more
complicated.

The profile likelihood function is an attempt to eliminate the
nuisance parameters from the likelihood function by replacing them
with their conditional maximum likelihood estimates (denoted, for
example, $\hat{\hat{{b}}}\ $).  The profile likelihood for
$L_P$ in Eq.~\ref{eq:lp} is given by
$L(x,y|\mu_0,\hat{\hat{b}}(\mu_0))$, with
\begin{eqnarray}
\hat{\hat{b}}(\mu_0) &=& \frac{x+y-(1+\tau)\mu_0}{2(1+\tau)}\\\nonumber
&+& \frac{\sqrt{(x+y-(1+\tau)\mu_0)^2+4(1+\tau)y\mu_0}}{2(1+\tau)}.
\end{eqnarray}
The relevant likelihood ratio is then
\begin{equation}
\lambda_P(\mu|x,y) = \frac{L(x,y|\mu_0,\hat{\hat{b}}(\mu_0))}{L(x,y|\hat{\mu},\hat{b})},
\end{equation}
where $\hat{\mu}$ and $\hat{b}$ are the unconditional maximum
likelihood estimates.

\begin{figure}
\epsfig{file=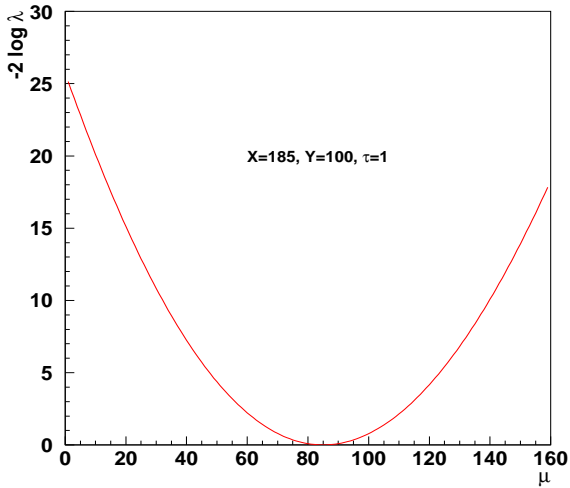, width=.45\textwidth, height=.4\textwidth}
\caption{The profile likelihood ratio $-2\ln \lambda$ versus the
signal strength $\mu$ for $y=100$, $\tau=1$, and $x=x_{crit}(y)=185$.}
\label{fig:profile_example}
\end{figure}

One of the standard results from statistics is that the distribution
of $-2\ln \lambda$ converges to the $\chi^2$ distribution with $k$
degrees of freedom, where $k$ is the number of parameters of interest.
In our example $k=1$, so a $5\sigma$ confidence interval is defined by
the set of $\mu$ with $-2\ln \lambda(\mu|x,y) < 25$.
Figure~\ref{fig:profile_example} shows the graph of $-2\ln
\lambda(\mu|x,y)$ for $y=100$ at the critical value of $x$ for a
$5\sigma$ discovery.

At PhyStat2003, Nancy Reid presented various adjustments and
improvements to the profile likelihood which speed asymptotic
convergence properties.\cite{Reid:2003vb} Cousins considers these
methods in more detail from a physicist
perspective.\cite{Cousins:phystat05}

Only recently was it generally appreciated that the method of {\sc
Minuit}\cite{Minuit} commonly used in HEP corresponds to the profile
likelihood intervals.  The coverage of these methods is not
guaranteed, but has been studied in simple
cases.\cite{Rolke:2000ij,Rolke:2004mj}  These likelihood-based
techniques are quite promising for searches at the LHC, but their
coverage properties must be assessed in the more complicated context
of the LHC with weighted events and several channels.  In particular,
the distribution of $q$ in Eq.~\ref{E:sb_exponentiation} is often
highly non-Gaussian.

\subsection{The Neyman Construction with Systematics}

Linnemann's preferred method, $Z_{Bi}$, is related to the familiar
result on the ratio of Poisson means.\cite{James:1980}  Unfortunately,
the form of $Z_{Bi}$ is tightly coupled to the form of
Eq.~\ref{eq:lp}, and can not be directly applied to the more
complicated cases described above.  However, the standard result on
the ratio of Poisson means\cite{James:1980} and Cousins'
improvement\cite{Cousins:1998} are actually special cases of the
Neyman construction with nuisance parameters (with and
without conditioning, respectively).

Of course, the Neyman construction does generalize to the more
complicated cases discussed above.  Two particular types of
constructions have been presented, both of which are related to the
profile likelihood ratio discussed in Kendall's chapter on likelihood
ratio tests \& test efficiency.\cite{Kendall}  This relationship often
leads to confusion with the profile likelihood intervals discussed in
Sec.~\ref{sec:likelihoodintervals}.

The first method is a full Neyman construction over both the
parameters of interest and the nuisance parameters, using the profile
likelihood ratio as an ordering rule.  Using this method, the nuisance
parameter is ``projected out'', leaving only an interval in the
parameters of interest.  I presented this method at PhyStat2003 in the
context of hypothesis testing,\footnote{In simple hypothesis testing
$\mu$ is not a continuous parameter, but only takes on the values
$\mu_0=0$ or $\mu_1=s$.} and similar work was presented by Punzi at
this conference.\cite{physics_03_10108,punzi:phystat2005} This method
provides coverage by construction, independent of the ordering rule
used.

The motivation for using the profile likelihood ratio as a test
statistic is twofold.  First, it is inspired by the Neyman-Pearson
lemma in the same way as the Feldman-Cousins ordering rule.  Secondly,
it is independent of the nuisance parameters; providing some hope of
obtaining similar tests.\footnote{Similar tests are those in which the
critical regions of size $\alpha$ are independent of the nuisance
parameters.  Similar tests do not exist in general.}  Both Punzi and
myself found a need to perform some ``clipping'' to the acceptance
regions to protect from irrelevant values of the nuisance parameters
spoiling the projection.  For this technique to be broadly applicable,
some generalization of this clipping procedure is needed and the
scalability with the number of parameters must be
addressed.\footnote{A Monte Carlo sampling of the nuisance parameter
space could be used to curb the curse of
dimensionality.\cite{physics_03_10108}}

The second method, presented by Feldman at the Fermilab conference in
2000, involves a Neyman construction over the parameters of interest,
but the nuisance parameters are fixed to the conditional maximum
likelihood estimate: a method I will call the {\it profile
construction}.  The profile construction is an approximation of the
full construction, that does not necessarily cover.  To the extent
that the use of the profile likelihood ratio as a test statistic
provides similar tests, the profile construction has good coverage
properties.  The main motivation for the profile construction is that
it scales well with the number of nuisance parameters and that the
``clipping'' is built in (only one value of the nuisance parameters is
considered).

It appears that the {\sc chooz} experiment actually performed both the
full construction (called ``FC correct syst.'')  and the profile
construction (called ``FC profile'') in order to compare with the
strong confidence technique.\cite{Nicolo:2002id}

Another perceived problem with the full construction is that bad
over-coverage can result from the projection onto the parameters of
interest.  It should be made very clear that the coverage probability
is a function of both the parameters of interest and the nuisance
parameters.  If the data are consistent with the null hypothesis for
{\it any} value of the nuisance parameters, then one should probably
not reject it.  This argument is stronger for nuisance parameters
directly related to the background hypothesis, and less strong for
those that account for instrumentation effects.  In fact, there is a
family of methods that lie between the full construction and the
profile construction.  Perhaps we should pursue a hybrid approach in
which the construction is formed for those parameters directly linked
to the background hypothesis, the additional nuisance parameters take
on their profile values, and the final interval is projected onto
the parameters of interest.

\section{Results with the Canonical Example}

\begin{figure}
\epsfig{file=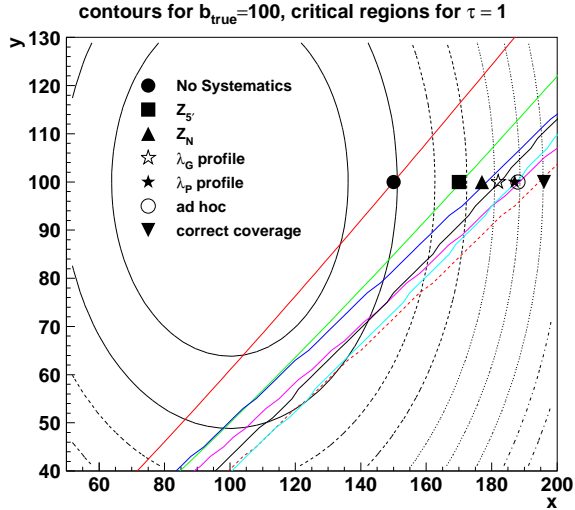, width=.45\textwidth}
\caption{A comparison of the various methods critical boundary
$x_{crit}(y)$ (see text).  The concentric ovals represent contours of
$L_G$ from Eq.~\ref{eq:lg}.}\label{fig:critregions}
\end{figure}

Consider the case $b_{true}=100$, $\tau=1$ ({\it i.e.} 10\% systematic
uncertainty).  For each of the methods we find the critical boundary,
$x_{crit}(y)$, which is necessary to reject the null hypothesis $\mu_0
= 0$ at $5\sigma$ when $y$ is measured in the auxiliary measurement.
Figure~\ref{fig:critregions} shows the contours of $L_G$, from
Eq.~\ref{eq:lg}, and the critical boundary for several methods.  The
far left curve shows the simple $s/\sqrt{b}$ curve neglecting
systematics.  The far right curve shows a critical region with the
correct coverage.  With the exception of the profile likelihood,
$\lambda_P$, all of the other methods lie between these two curves
({\it ie.} all of them under-cover). The rate of Type~I error for
these methods was evaluated for $L_G$ and $L_P$ and presented in
Table~\ref{tab:coverage}.

The result of the full Neyman construction and the profile
construction are not presented.  The full Neyman construction covers
by construction, and it was previously demonstrated for a similar case
($b=100$, $\tau=4$) that the profile construction gives similar
results.\cite{physics_03_10108} Furthermore, if the $\lambda_P$ were
used as an ordering rule in the full construction, the critical region
for $b=100$ would be identical to the curve labeled ``$\lambda_P$
profile'' (since $\lambda_P$ actually covers).  

It should be noted that if one knows the likelihood is given by
$L_G(x,y|\mu,b)$, then one should use the corresponding profile
likelihood ratio, $\lambda_G(x,y|\mu)$, for the hypothesis test.
However, knowledge of the correct likelihood is not always available
(Sinervo's Class II systematic), so it is informative to check the
coverage of tests based on both $\lambda_G(x,y|\mu)$ and
$\lambda_P(x,y|\mu)$ by generating Monte Carlo according to
$L_G(x,y|\mu,b)$ and $L_P(x,y|\mu,b)$.  In a similar way, this
decoupling of true likelihood and the assumed likelihood (used to find
the critical region) can break the ``guaranteed'' coverage of the
Neyman construction.

It is quite significant that the $Z_N$ method under-covers,
since it is so commonly used in HEP.  The degree to which the method
under-covers depends on the truncation of the Gaussian posterior
$P(b|y)$.  Linnemann's table also shows significant under-coverage
(over estimate of the significance $Z$).  In order to obtain a
critical region with the correct coverage, the author modified the
region $x_{crit}(y) = x_{crit}^{Z_N}(y) + C$ and found $C=16$ provided
the correct coverage.  A discrepancy of 16 events is not trivial!

\begin{tablehere}
\caption{Rate of Type I error interpreted as equivalent $Z\sigma$ for
various methods designed for a $5\sigma$ test.  Monte Carlo events are
generated via either $L_G$ or $L_P$.  The critical $x$ for $y=100$ is
also listed for easy comparison.}\label{tab:coverage}
\begin{tabular}{|l|c|c|c|}
\hline
Method  & $L_G$ ($Z\sigma$) & $L_P$ ($Z\sigma$) & $x_{crit}(y=100)$  \\ \hline
No Syst             & 3.0 & 3.1 & 150 \\
$Z_{5'}$            & 4.1 & 4.1 & 171\\
$Z_N$ (Sec.~\ref{sec:CousinsHighland})   & 4.2 & 4.2 & 178\\
{\it ad hoc}        & 4.6 & 4.7 & 188\\
$Z_\Gamma = Z_{Bi}$ & 4.9 & 5.0 & 185\\
profile $\lambda_P$ & 5.0 & 5.0 & 185\\ 
profile $\lambda_G$ & 4.7 & 4.7 & $\sim$182\,\,\,\, \\\hline
\end{tabular}
\end{tablehere}

Notice that for large $x,y$ the Bayesian-frequentist hybrid $Z_N$
approaches $Z_{5'}$, where the the critical region is of the form
$x_{crit}(y) = y/\tau + n\sqrt{y/\tau}$.  Because the boundary is very
nearly linear around $y_0$, one can find the value of $n$ that gives
the proper coverage with a little geometry.  In particular, the number
$n$ needed to get a $Z\sigma$ test gives
\begin{equation}
x_{crit}(y) = y/\tau + Z\sqrt{1+1/\tau m^2}\sqrt{y/\tau}
\end{equation}
where
\begin{equation}
m = \left( 1+\frac{Z}{2\sqrt{y/\tau}}\right)^{-1}
\end{equation}
The $m^2$ factor can be seen as a correction to the $Z_{5'}$ and $Z_N$
results.  Notice that the correction is larger for
higher significance tests.  As an {\it ad hoc} method, I experimented
with the $Z_N$ method replacing $\tau$ with $\tau m^2$ in
the posterior $P(b|y)$. The coverage of this {\it ad hoc} method is
better than $Z_N$, but not exact because $x,y$ are not sufficiently
large.

\section{Conclusions}

I have presented the statistical challenges of searches at the LHC and
the current state of the statistical methods commonly used in HEP.  I
have attempted to accurately portray the complexity of the searches,
explain their key ingredients, and provide a practical example for
future studies.  Three classes of methods, which are able to
incorporate all the ingredients, have been identified: hybrid
Bayesian-frequentist methods, methods based on the Likelihood
Principle, and frequentist methods based on the Neyman construction.

The Bayesian-frequentist hybrid method, $Z_N$, shows significant
under-coverage in the toy example considered when pushed to the
$5\sigma$ regime.  While Bayesian might not care about coverage,
significant under-coverage is undesirable in HEP. Further study is
needed to determine if a more careful choice of prior distributions
can remedy this situation -- especially in more complex situations.
The improved coverage of $Z_\Gamma$ may give some guidance.

The methods based on the likelihood principle have gained a great deal
of attention from HEP in recent years.  While the methods appear to do
well in the toy example, it requires further study to determine their
properties in the more realistic situation with weighted events.

Slowly, the HEP community is coming to grips with how to incorporate
nuisance parameters into the Neyman construction.  Several ideas for
reducing the over-coverage induced by projecting out the nuisance
parameters and reducing the computational burden have been presented.
A hybrid approach between the full construction and the profile
construction should be investigated in more detail.

Finally, it seems that the HEP community is approaching a point where
we appreciate the fundamental statistical issues, the limitations of
some methods, and the benefits of others.  Clearly, the philosophical
debate has not ended, but there seems to be more emphasis on practical
solutions to our very challenging problems.

\section*{Acknowledgments}

I would like to thank the many people that helped in preparing this
review.  In particular, Bob Cousins, Jim Linnemann, Gary Feldman, Jan
Conrad, Fredrik Tegenfeldt, Wolfgang Rolke, Nancy Reid, Gary Hill, and
Stathes Paganis.  I would also like to thank Louis Lyons for his
continuing advice and the invitation to speak at such an enjoyable and
productive conference.

This manuscript has been authored by Brookhaven Science Associates
under Contract No. DE-AC02-98CH1-886 with the U.S. DOE.
  The U.S. Government retains, and the publisher, by accepting
the article for publication, acknowledges, a world-wide license to
publish or reproduce the published form of this manuscript, or allow
others to do so, for the U.S. Government purposes.

\bibliographystyle{ws-procs11x85}
\bibliography{Cranmer_stats}

%
%
%
%
%

\end{document}